# Integration of CAD and rapid manufacturing for sand casting optimisation

*Alain Bernard,*
*Jean-Charles Delplace,*
*Nicolas Perry and*
*Serge Gabriel*

### The authors

**Alain Bernard** is Professor in IRCCyN research laboratory (Industrial engineering group) and in Ecole Centrale de Nantes (France).
**Jean-Charles Delplace** is the PhD student that conducted this research work.
**Nicolas Perry** is Assistant Professor in IRCCyN research laboratory (Industrial engineering group) and in Ecole Centrale de Nantes (France).
**Serge Gabriel** is the head of SMC Colombier-Fontaine company and also the head of R&D of AFE group.

### Abstract

In order to reduce the time and the costs of the products development in the sand casting process, the SMC Colombier Fontaine company has carried out a study based on a tooling manufacturing with a new rapid prototyping process. This evolution allowed the adequacy of the geometry used for the simulation to the tooling employed physically in the production. This allowed a reduction of the wall thickness to 4mm and retains manufacturing process reliable.

## Introduction

SMC COLOMBIER FONTAINE is a company of the AFE METAL group, which uses a sand casting process to manufacture steel primary parts. To reduce the « time to market », the primary part producers also need to reduce the industrialisation time and costs. These obligations and the intention of improving the process performances brought SMC to develop the numerical technologies to manufacture the tooling.

In order to validate the parts design, the company uses layer manufacturing technologies to make geometrical prototypes. After it has been validated, it can serve to get functional prototypes with a lost pattern process. Both are used to allow the entire group involved in the project to validate the solutions, by the application of the concurrent engineering concepts for rapid product development [Bernard, 1999 ; Bernard et al., 2000 ; Xue, 1996].

These improvements are nowadays inked into the company culture. The next steps for the industrialisation time and costs reduction is the introduction of the rapid tooling technologies. For the sand casting, the processes are the Stereolithography, the Powder melting (EOS-DTM), the layer cutting out (LOM), the layer machining (Stratoconception), and the 3D machining, which is the most used [Barlier, 1998].

But to be able to generalise these kinds of technologies, the CAD study is an inescapable step in the part industrialisation. The whole improvements led by the numerical technologies should justify the investment in time spent to do CAD modelling of all the different elements in the sand casting process. But, like most companies that manage a large product mix, SMC uses at the moment CAD only for part design.

The content of this paper is to demonstrate the efficiency of the proposed approach which objective is to generalise a complete numerical reference by introducing CAD, simulation and new manufacturing technologies in order to be able to optimise the product and its process.

## Industrial context of the project

SMC is characterised by a strong product mix with 7 different business lines, about 1000 revolving references per year and an average of 15 new tooling per month. Furthermore, the sand casting process needs, at the moment from 20 to 40 hours to complete a CAD study; that is to say, model the part, the master pattern, the pattern-plates, the cluster and simulate the fill up and the solidification. To generalise this kind of study to all the parts is not possible at the moment compared to the technical office capacities. Therefore, the toolmaker suppliers use the traditional moulding techniques to manufacture the pattern plates from the 2D drawing or from the part CAD model. The random repeatability of the tooling penalises the simulation accuracy. Therefore, there are only some clusters that are simulated and the results are interpreted with cautiousness.

The numerical tooling should lead to improvements from the manufacture accuracy. Indeed, at the moment, the traditional moulding technique accuracy is about 1 millimetre when the rapid tooling technologies are closer to a tenth. It allows the parts tolerances improvement, the machining allowances and the flash reduction. It also gets more reliable simulation results by the adequacy between the CAD model and the real cluster geometry. It optimises the overall product industrialisation time and costs by the reduction of the iterations needed to perfecting the serial production design. Furthermore, the improvements in the metallurgical quality allow the reduction of the wall thickness while bringing down the misrun or the breaking risks. Therefore, the casting parts become a technological alternative less and less risky compared to the welded or the machined parts.

**Content of the study**

To be able to validate these concepts, SMC made the following study. The part chosen for the trial is manufactured, at the moment, by a welding process. To modify the process, the most important criteria are price, weight, mechanical properties and metallurgical qualities. To respect the price and the weight requirements, SMC was obliged to reduce the wall thickness under the limits of the process, that is to say 4 mm for a minimum of 6 mm. The study started with the part design and their assembling to create the cluster (Figure 1).

**Figure 1** Cluster CAD model

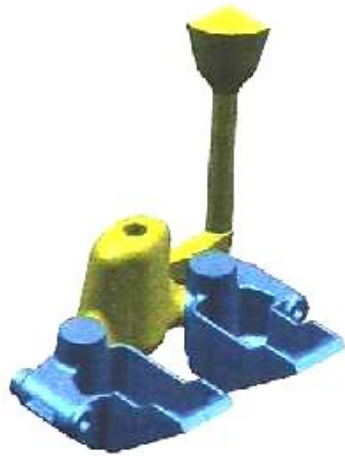

Thanks to this model, the filling and the solidification simulation brought to the fore several risk areas. The CAD model was corrected and the simulation validated the modifications (figure 2).

**Figure 2** Filling up and solidification simulations

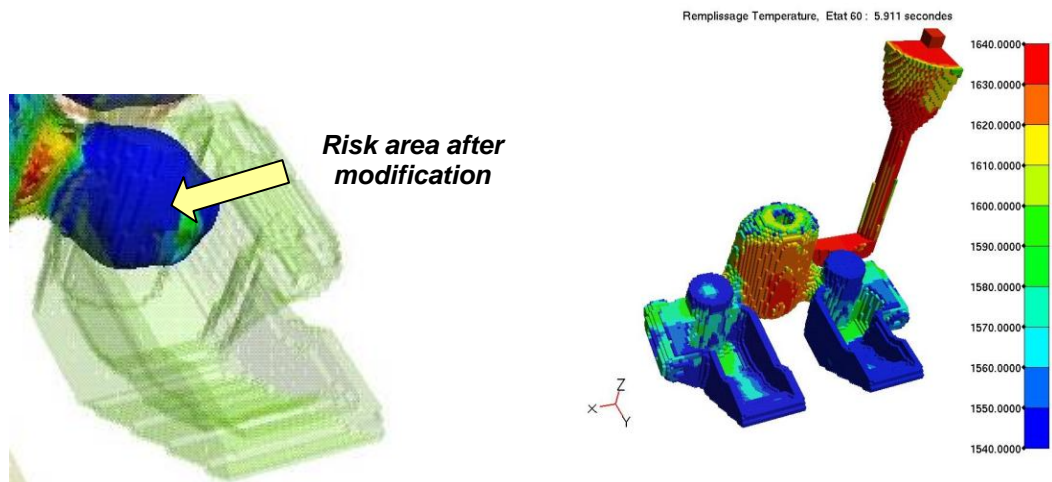

*Risk area after modification*

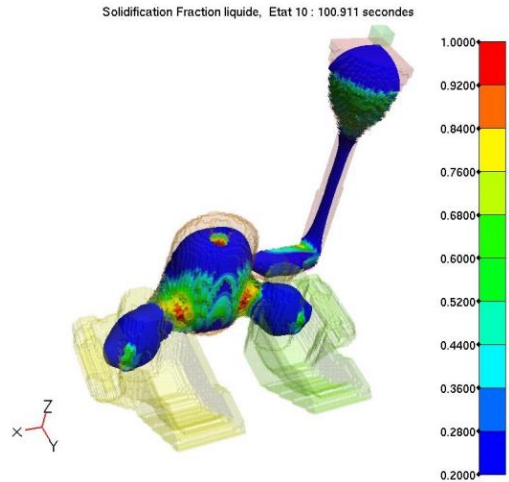

The pattern plates and the core boxes were designed with the cluster modelling. A new rapid tooling process manufactured the master-patterns. This technology was introduced in 1998 by OPTOFORM (Schaeffer and al., 1998], which was purchased in 2001 by 3D Systems [3D Systems, 2001]. This validation for the sand casting process played a part in the industrial development made jointly with DASSAULT AVIATION and AEDS CCR. This process, close to the Stereolithography, brings into play materials exploitable in paste form, which allows a large application range. Indeed, the resin paste permits a high level of additional material, which increases the mechanical properties. The paste is set down into thin layers with specific scrapers, then, solidified by a laser. This process also uses a ceramic paste to obtain parts or cores [Doreau, 2001]. At present, the interest of such technology has been validated on a technical point of view concerning the manufacturing of ceramic components and tools but also for alumna and metallic parts. Unfortunately, it is not definitively assumed that layered-manufacturing techniques is economically comparable with machining for the manufacturing of the core-boxes.

**Figure 3** OPTOFORM process diagram

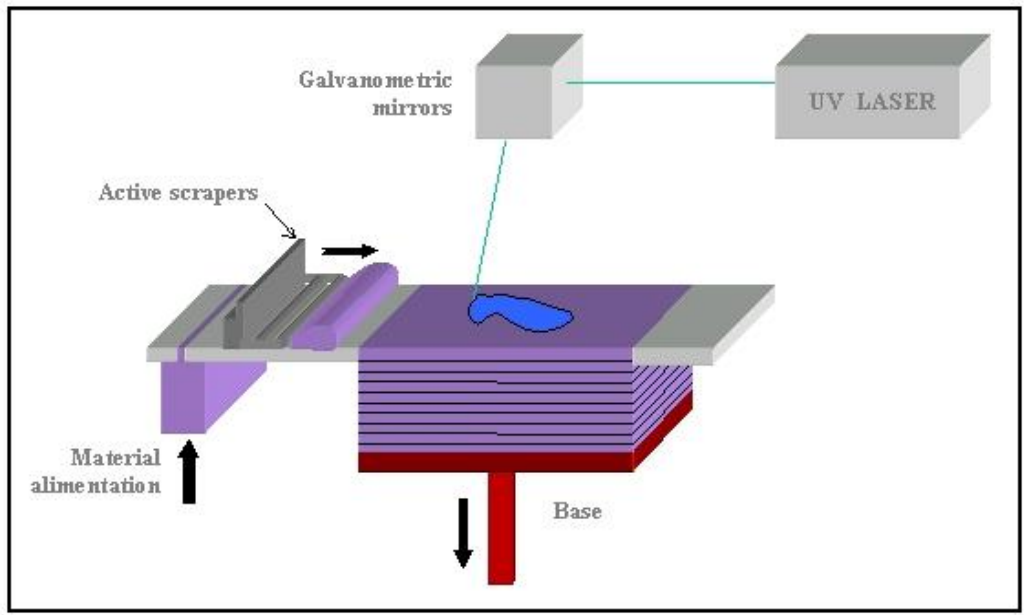

.

The development machine capacities were 250x250x250mm that obliged us to manufacture the patterns in two parts. The overall manufacture time was 20 hours. The profitability of the use of such technology has been demonstrated for complex parts. But for simple shape tools, machining and manual finishing allow obtaining a better economical ratio. (figure 4).

**Figure 4** Pattern and core boxes rough prints

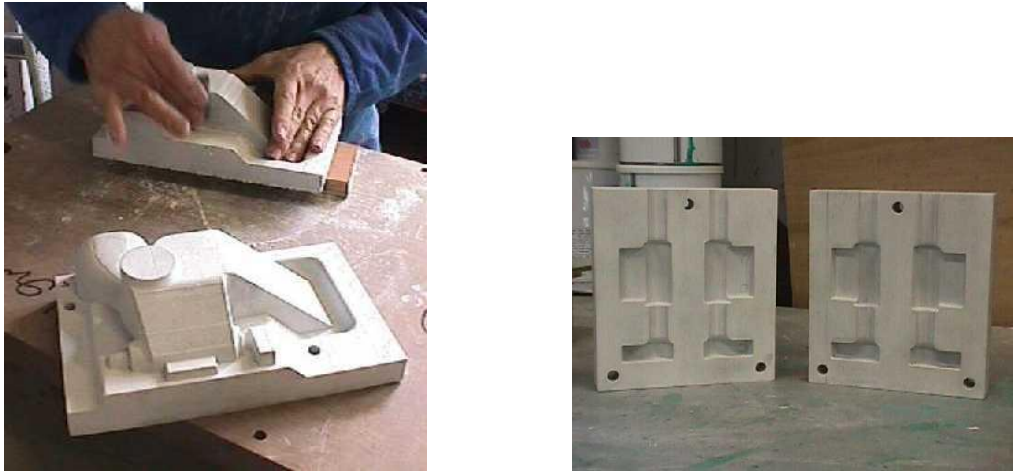

Afterwards, the pattern was inserted into the standard pattern plate. 20 hours of work were needed for the assembly. The manufacture and the assembly time can be optimised easily. Indeed, the marketed machine capacities are now of 500x500x500mm at the maximum, which means that the pattern should be manufactured in one time. Furthermore, the holes necessary to assemble the centring lugs and the whistler-spiracle should be inserted into the numerical model to avoid the drilling. The tooling has finally received a surface treatment to improving the abrasion resistance. (figure 5).

**Figure 5** Tooling assembly and pattern plate

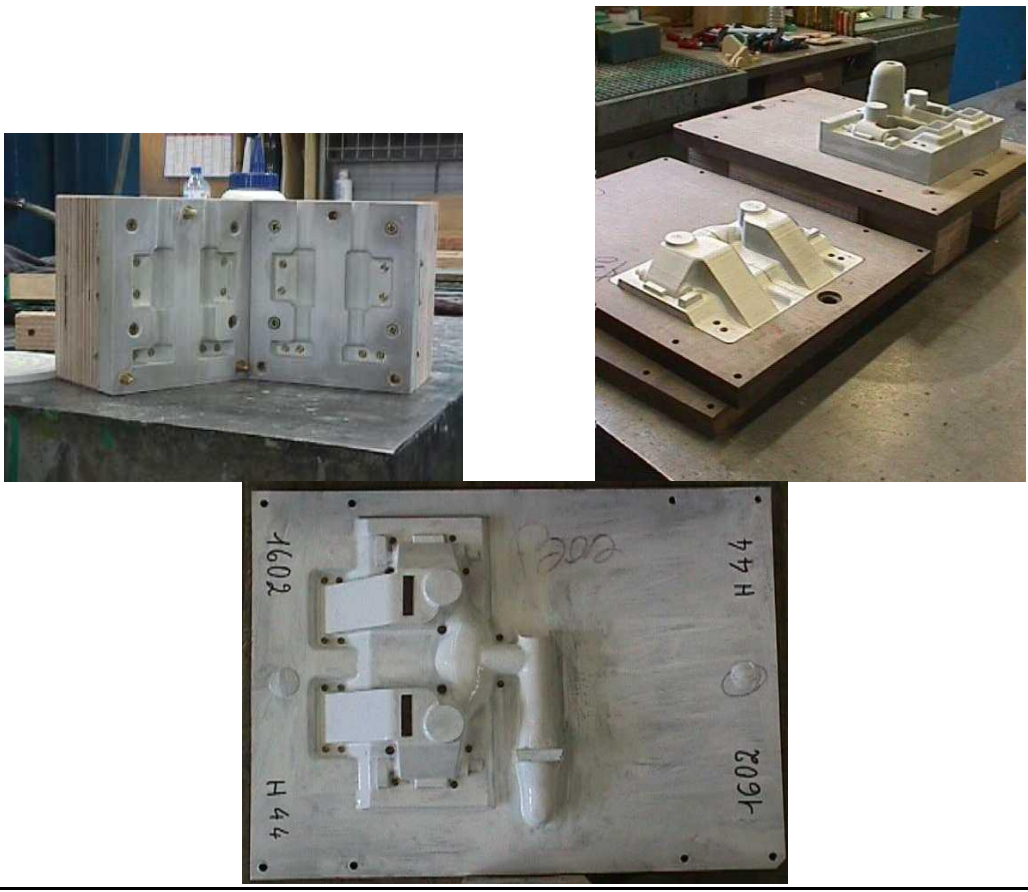

### Technical results

500 moulds were produced and the results were very good. The core box and the pattern plate did not show any defects. SMC uses a fully automatic molding site that uses natural set sand (without resin polymerization). It is plastered against the pattern plate by an explosion and then some piston squeeze the sand mould with a pressure of 12 Kg/cm² or 84 bar. Therefore, the molding process used in SMC is very aggressive in terms of abrasion and compression resistance. This operation takes place in the ambient temperature and the tools do not warm up a lot. So, with a maximum of 40°C, the temperature resistance is not as important for example as it is for the plastic injection. The pattern-plate was surface treated and, in the most exposed areas, the protection coat disappeared but the rough material has not suffered from the abrasion (Figure 6).

**Figure 6** Test closing mould

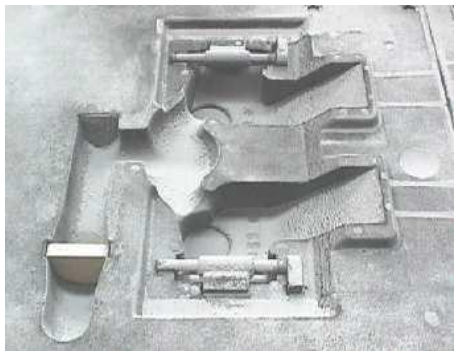

The OPTOFORM process was validated, without any problems, with this small production. Unfortunately, we did not have the opportunity to test it for a bigger production, but the tools state at the end of the trial, looked like the production ones. Therefore, the estimations about the tooling production capacity are about several thousands of moulds (figure 7).

**Figure 7** Final part

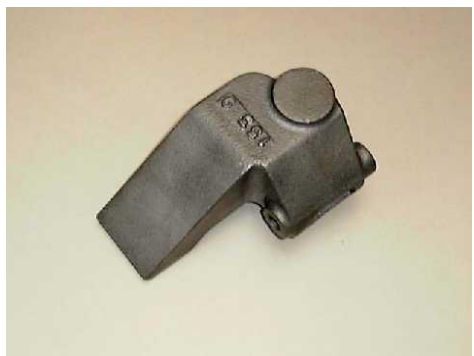

Furthermore, the geometrical requirements have been respected (JS12) and allowed the adequacy between the physical cluster and the virtual one used for the simulation. Indeed, the metallurgical analyse of the parts and of the runners shown results very close to the simulation. because the tooling manufacturing took place during the OPTOFORM process development, manufacturing and assembling were not optimised, which makes it difficult to report the exact figures of the different improvements.

### Time and cost progresses

The following figures show the average time of all the operations that composed the industrialisation with and without the numerical channel. The improvement in term of time is about 20%. Unfortunately, this trial has shown the difficulty to justify the development costs payback. The following figure shows the overall development costs for the trial (figure 8).

**Figure 8** Product development time and cost

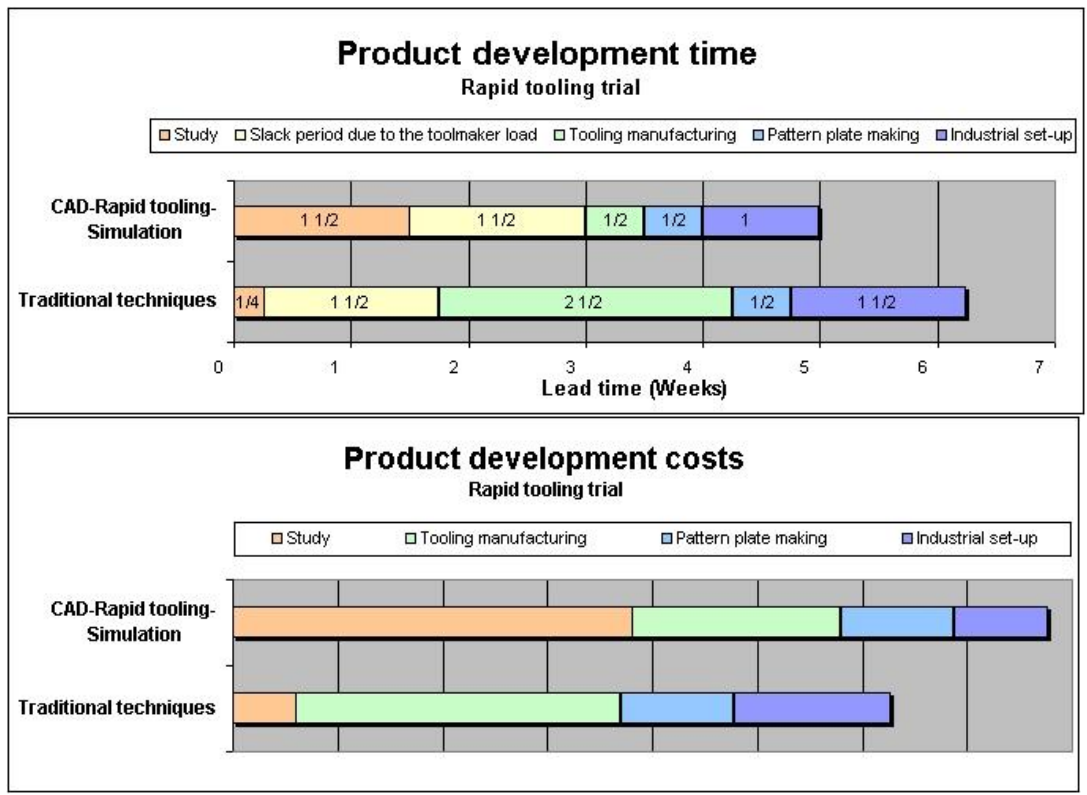

The overall development costs were about 15% more with the numerical channel when with the traditional techniques. This difference was due to the difficulty to realise the complex moulding study with the CAD tools; that is to say, model the complex split joints, manage the moulding looses, etc. At the moment, the toolmaker has progressed to improve this skill while the casting designer has lost it. Indeed, when the CAD was introduced in the foundry, the choice was restricted to the professional systems used by the customers (CATIA, Pro-engineer, etc.). Actually, the foundry design offices are composed of casting engineer without CAD skill. The training and the integration of this kind of tools are difficult. Furthermore, the hourly rate is expensive, the time spent is important and the design costs rocket. These are the reasons why the payback was not possible. Therefore, the way to generalise the CAD study is to pass by the user-friendly improvements and the specifics tools developments.

SMC started this mutation during 2000 by the introduction of user-friendly CAD software. The main three are Think3, SolidEdge and SolidWorks, which are all modern software, easy to use and to set-up on a usual PC. The choice was SolidWorks, which has the best user-friendliness and the most compatibility with the window products. It introduces a new outlook on the CAD. It stops thinking of it as a professional tool and thinks of it as an office automation application. The CAD software becomes a usual icon in the office-shortcut bar like Word, Excel or PowerPoint. Furthermore, the foundry designers are casting and not CAD professional. So, they need tools to help them to design to parts with the easier and the quicker way to do it.

Furthermore, SolidWorks was installed on laptops, which gives the relevant flexibility to train and use the software. It also leads improvements in the ways to communicate with the customer. The concurrent engineering becomes as easy as to go to the customer and to work with him straightaway on the CAD project with the laptop. It also simplifies the way to send the files by e-mail for example to a prototype supplier or to a representative. These improvements are common nowadays but they took much more time with the CAD systems that were installed on workstations.

The flexibility introduced by the couple SolidWorks-laptops leads to a massive reduction of the CAD hourly rate due to the system price and to the time spend on it. The next figure shows the cost results for the complex parts in 2001 (Figure 9)

**Figure 9** New product development costs

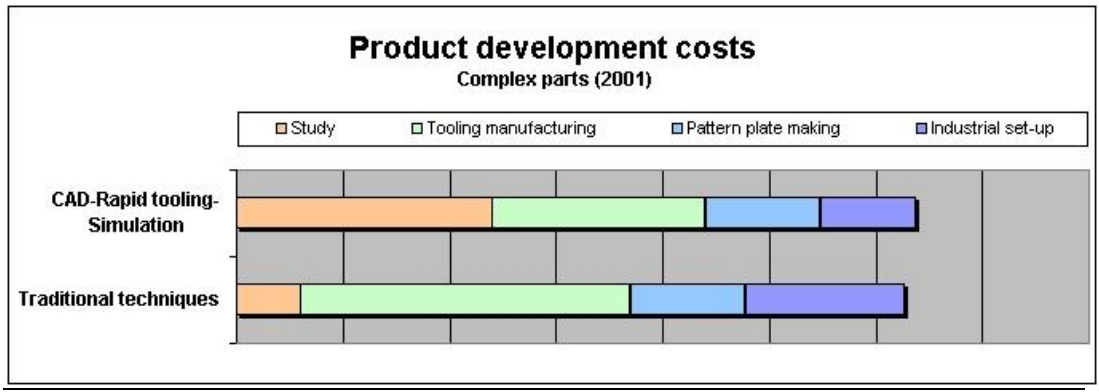

For the parts with cores, the design costs are getting closer but the paybacks are not relevant enough yet to justify the design process modification. But for the parts without core, the easy way to use the CAD tools brought SMC to generalise the rapid tooling naturally. The following figure shows the costs to manufacture the simple parts tools (Figure 10).

**Figure 10** Cost corresponding to sample parts

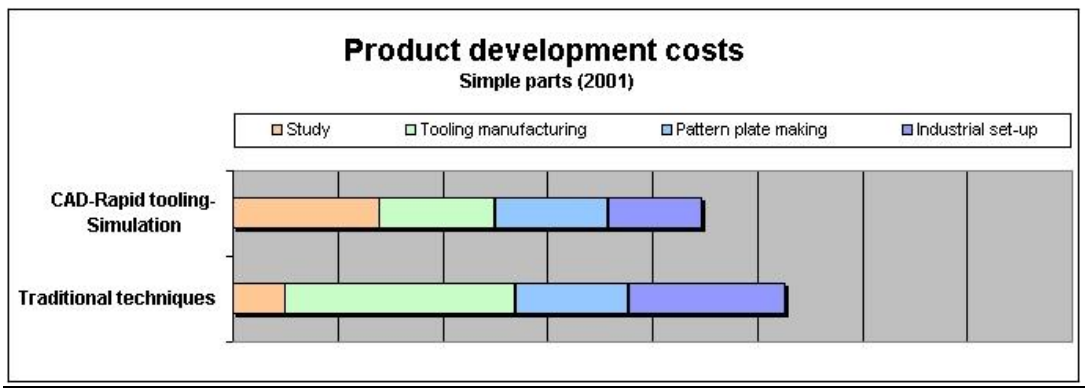

It highlights the reduction of the design costs thanks to a difference about 20% cheaper with the numerical channel when with the traditional techniques. These results can be more improved by the development of specific design tools. They should manage the quicker way to complete moulding study from part design. In this way, the numerical chain has been generalised for the complex parts as well.

## Conclusions

By the achievement of this study, SMC introduces a rapid prototyping technology able to manufacture the production tooling for the sand casting process. The 500 moulds manufactured prove the feasibility for a small production and the good tooling statement allows expecting the same results for the medium production (several thousands). It also proves the improvement in the simulation reliability thanks to the good metallurgical results, in spite of wall thickness reduction under the process limits. Next to the technical validation, the payback analyses highlight the problem of cost lead by the achievement of the entire CAD moulding study. The successful introduction of SolidWorks reduces part design time and CAD hourly rate. In order to reduce the design costs sufficiently to generalise the numerical chain, SMC decided a specific development of particular design tools completely dedicated to the design of raw parts and tools. The development has been achieved at the end of 2002 [Bernard et al, 2002a ; Bernard et al, 2002b] and, since this date, all the new products have been designed and industrialised using the numerical chain and the design methodology [Bernard et al, 2003]. All these improvements have shown SMC an interesting way of progress to bring the sand casting offer to be more competitive.